# Countering the Forgetting of Novel Health Information with 'Social Boosting'

**Vaibhav Krishna[*] PhD**

Yale Institute of Network Science,
Yale University,
Suite 393A, 17 Hillhouse Avenue,
New Haven, CT 06511, USA
vaibhav.krishna@yale.edu

**Prof. Nicholas A. Christakis MD**

Yale Institute of Network Science,
Yale University,
Suite 393A, 17 Hillhouse Avenue,
New Haven, CT 06511, USA

*corresponding author

# Countering the Forgetting of Novel Health Information with 'Social Boosting'

Vaibhav Krishna[1*], Nicholas A. Christakis[1]

ABSTRACT

**Objective:** To assess long-term knowledge retention of a sustained educational intervention regarding maternal and child health.

**Setting:** A multi-arm cluster-randomized controlled trial in 110 villages in Honduras' Copan region, involving 3910 targeted individuals in 2038 households from October 2015 to December 2019.

**Participants:** Residents aged 12 and older were eligible for a photographic census which was completed with 12353 and 8250 at baseline and endline surveys, respectively, constituting over 81·2% of the population.

**Intervention:** 22-month household-based counseling intervention aiming to improve knowledge, attitudes, and practices related to maternal, neonatal, and child health.

**Primary outcome measures:** Primary outcomes were knowledge retention at the endline survey related to prenatal/postnatal care behaviors, breastfeeding, treatment of diarrhea and respiratory illness, and reproductive health and norms.

**Results:** We found that, compared to targeted individuals with none or no more than two friendship ties, those with a higher number of friendship ties exhibited a substantially greater probability of correct response after the knowledge intervention. Each additional friendship ties was associated with a 27·8% (95% CI 13·2%-44·4%, $p < 0·000$) increase in the likelihood of correctly retaining the knowledge about a previously unheard riddle related to proper cord care in newborns; a 82% (95% CI 42·2%-136·6%, $p < 0·000$) increase in identifying that women should start going for prenatal checkups at <12 weeks during pregnancy; a 29·4% (95% CI 11·9%-50·7%, $p=0·004$) increase in knowing that newborn should only be given breastmilk until 6 months; and a 22·2% (95% CI 7·5%-38·8%, $p=0·015$) increase in identifying that pneumonia is a postnatal danger sign.

**Conclusion:** The number of friends that targeted individuals have – a proxy for increased opportunities for social interactions – has a significant association with their likelihood of retaining accurate knowledge across multiple outcomes at an endline survey related to prenatal, maternal, and newborn care.

*corresponding author

[1] Yale Institute of Network Science, Yale University, New Haven, Connecticut, USA

## Research in context

### Evidence before this study

The prevalence of false or misleading information, especially with respect to health care practices, poses a threat. While various intervention strategies have shown promise for spreading correct knowledge in a population, the effectiveness of such interventions can decay. We did a PubMed and Google Scholar search for studies reporting the efficacy of intervention approaches that used debunking or "pre-bunking" published up to March 31, 2025, with the search terms "misinformation", 'infodemics", "interventions", or 'long term effectiveness". We identified two studies that suggested the effectiveness of a social booster intervention strategy that helped sustain the duration of learning after an educational intervention. However, implementation of regular such booster interventions would require substantial resources. We found no studies that evaluated the role of the detailed social structure in reinforcing the health knowledge post intervention.

### Added value of this study

We examined how traditional intervention strategies might benefit from leveraging social network interactions as a mechanism for reinforcement, which we refer to as "social boosting", to strengthen the retention of correct knowledge. We focused on a critically important topic, information about maternal and child health care. We found that social structure of local villages within which the intervened individuals live provides opportunities for the targeted individuals to discuss and internalize new knowledge.

### Implications of all the available evidence

Our results indicate that well-connected individuals within a social network experience an enhanced effectiveness of knowledge interventions. These individuals, who have increased opportunities for social interaction where they could engage in discussing any newly acquired knowledge with others, may be more likely to internalize and retain the information. These findings underscore the role of social interactions in reinforcing health knowledge interventions over the long term that could guide future designs of health policy or civil society work.

**INTRODUCTION**

Despite the importance of people being well-informed across many domains, repeated exposure to low-quality or false information can lead to its acceptance, even if it is harmful—a phenomenon known as the illusion of truth[1,2,3]. Examples range from widespread doubt-sowing messages on global warming that affect the progress of policymakers,[4] to viral conspiracy documentaries around the pandemic that have been linked to vaccine hesitancy,[5] to the disinformation related to new technologies like 5G resulting in vandalism[6].

To mitigate such adverse effects, researchers have proposed various intervention techniques – for instance, debunking[7], through post-hoc correction, or pre-bunking, based on inoculation theory, which serves both as a preemptive and therapeutic measure to enhance cognitive resilience[8,9]. While these intervention techniques have demonstrated efficacy, recent studies have highlighted a decay process[10,11] consistent with Ebbinghaus' well known forgetting curves[12]. Though a general consensus on the duration of the intervention effectiveness varies, recent longitudinal studies have often reported a complete decay of many knowledge interventions after just a few weeks[10,11,13]. As a result, even after corrections are made, individuals may persist in believing false information due to the continued influence effect (CIE)[14,15].

To address this decay, studies have suggested "booster" treatments that exploit insights into memory and the importance of rehearsal[16,17]. Drawing on the principles of biomedical inoculation, it is theorized that, analogous to vaccines that strengthen the biological immune system, for the "cognitive immune system," regular information boosters may be needed[18]. Thus, while some longitudinal studies have reported a decay starting after two weeks, studies administering booster messages have shown the effectiveness of up to at least 44 days, with a general consensus that boosters work when administered in the right form at the right time[11]. However, despite their effectiveness, the implementation of regular booster interventions would require substantial resources, posing significant scaling challenges.

In this study, we hypothesize that traditional intervention strategies can involve leveraging social network interactions as a mechanism for reinforcement to strengthen the retention of correct knowledge. Individuals who receive specific knowledge can internalize and consolidate this information by engaging in social interactions where, for instance, they at least had an *opportunity* to discuss it with others in the process. According to the cognitive theory of learning, generative processing occurs when individuals explain knowledge, facilitating deeper understanding[19]. Therefore, the act of explaining information to others (knowledge sharing) promotes deeper cognitive processing and elaborative encoding, which ultimately enhances memory retention [20,21].

Using data from a randomized controlled trial in which individuals received counseling interventions related to pregnancy and early infancy, and in which knowledge outcomes were assessed approximately 22 months after the intervention commenced, we examine how the targeted individuals' positions within the social network—specifically, the number of friends they interact

with—influence the long-term retention of accurate knowledge through what we term a "social booster treatment."

## METHODS

### Study design and population

Our primary analyses uses individual-level data from a randomized-controlled trial (RCT) study[22,23] conducted in the Department of Copan in western Honduras, an area characterized by high rates of neonatal and maternal morbidity and mortality. The parent trial evaluated a novel social network targeting technique in 176 villages in the Copan region; of these, 110 were included in the current assessment (specifically, we included only those villages in which participants were assigned to treatment based on randomized target selection within the villages or villages in which either none or all the households received treatment). All individuals living in the study village and aged 12 years or above were eligible to participate. All the participants provided informed consent before enrollment in the study and the Yale IRB and the Honduran Ministry of Health approved all the data collection procedures (Protocol #1506016012).

A total of 24,702 individuals living in 10,013 households in 176 villages (constituting 81·2% of the total population) were enrolled and participated in the baseline social network and attributes survey. The targeted households and individuals within these villages received a 22-month educational intervention (between Nov 2016 to Aug 2018, involving monthly, in-home visits) and were surveyed periodically. We rely on survey data collected during two waves – wave 1 [Oct 2015 – Jun 2016; (W1)] and wave 3 [Jan 2019 – Dec 2019; (W3)]. Social networks within each village were mapped using Trellis software[24] and various "name-generator" questions were used to capture different social relationships.

The primary relationships we examined here used three key name generators based on the individual subjects identified as those with whom they (i) "spent free time", (ii) discussed "personal or private" matters, or (iii) were "close friends."  The social network thus obtained were symmetrized (i.e., we counted a tie either when an ego nominated an alter, or vice versa). In addition, as we are interested in capturing knowledge sharing through social interactions, relationships based on another key name generator were included for additional analysis, namely the people an ego identified as those to whom they gave or from whom they got health advice.

A two-stage factorial design was used in the parent trial; in the first stage, villages were randomly assigned a dosage level, or the proportion of households targeted for intervention per village (0, 0·05, 0·1, 0·2, 0·3, 0·5, 0·75, 1); and, in the second stage, households were randomly assigned to the intervention. The current randomized assessment focuses on the 88 villages of which 66 villages were assigned to a proportion of randomly targeted households greater than 0 and less than 1 in a village, and 22 villages where all the households received the intervention. In addition, 22 further villages where no household received the intervention—i.e., villages with a dosage proportion equal to 0—were included and served as a control group for certain analyses.

The intervention assessment reported here establishes the correct knowledge response among the targeted individuals in the 88 villages at the endline survey (W3) collected 2 years after the baseline survey (W1). To identify key knowledge outcomes for evaluation, we compared responses at the endline survey for participants in the 22 fully treated villages (where everyone got the educational intervention) to the responses at the baseline survey. This comparison enabled the shortlisting of outcomes that demonstrated meaningful improvements in knowledge following the intervention, for use in the analysis in all the focal 88 villages.

While both the participants and the Community Change Agents (CCAs) who delivered the intervention were blinded to the criteria used for selecting the intervention households, complete masking of participants and CCAs was not feasible due to the nature of the intervention; however, ascertainment of outcomes in all villages was blinded as to treatment assignment.

**Procedures**

The RCT in this study was conducted with extensive local engagement, involving introducing the project to village leaders, securing local approvals, and managing local implementation of the study and intervention. CCAs, who were trained, compensated, and managed by World Vision, delivered the intervention – named *Proyecto Redes: Con Amor y Cuidados Madres y Bebés Sanos* (With Love and Care, Healthy Moms and Babies)[25] – across two years (November 2016–August 2018). Each targeted household received up to 22 sessions, typically lasting 1–2 hours and delivered monthly. These sessions covered 15 thematic modules based on a modified version of the globally recognized Timed and Targeted Counselling (ttC) framework[26].

The interventions were tailored according to the family's current circumstances and, accordingly, CCAs spoke to families regarding several health topics during every house visit. The educational content was designed to enhance knowledge about critical health topics including pregnancy preparedness, safe childbirth practices, newborn care, maternal nutrition, diarrheal and respiratory illness, and early childhood development. Furthermore, content delivery incorporated a behavior-change communication strategy based on the P Process tool, which uses narrative and negotiation techniques to foster engagement and agreement on health-promoting behaviors[27]. An important part of the visits involved designing sessions that included discussing relevant regional practices that are potentially dangerous to newborns. The modules were thus tailored to correct these prevailing health malpractices and misbeliefs.

CCAs used tablets during the visits, which supported standardized delivery through multimedia aids—where families access educational materials through videos, songs, and riddles—that facilitate better understanding and retention as well as serve as a data collection tool. Importantly, unlike many community-based health interventions, this program was not limited to expectant mothers but included households irrespective of composition, allowing a broader diffusion of knowledge within social networks. The intervention also did not rely on mass media or community mobilize groups, focusing instead on intensive, individualized home visits.

Two different and independent teams of people delivered the interventions and conducted the outcomes assessment. The survey instrument was designed to capture target outcomes and demographic information. Baseline data, including a photographic census and network mapping, were collected from June 2015 to June 2016 using the open-source Trellis platform[24]. Endline data collection, including a follow-up census and survey, occurred from January 2019 to December 2019 (W3)—approximately 4 months after the conclusion of the intervention.

**Outcomes**

The primary outcomes are responses to knowledge queries at the endline survey (W3) related to maternal and neo-natal healthcare practices, which can be grouped into the following: (1) prenatal care and pregnancy danger signs; (2) care of the mother during childbirth and the postpartum period; (3) newborn care and danger signs; (4) infant care (caring for children 1-6 months); (4) danger signs and seeking medical attention for acute respiratory infections and diarrheal illnesses; (5) reproductive health including life plan and importance of preventing teen pregnancy; and (6) importance of folic acid for mother and baby. All outcomes were measured via a standard endline survey instrument. Participants' responses were recorded for a total of 60 knowledge outcomes, each mapped to one or more of the 15 counseling modules delivered during the intervention.

To shortlist the key outcomes for analysis in the focal 88 villages, we assessed the improvement in knowledge response at the endline survey (W3) in the 22 villages where 100% of the residents got the intervention compared to responses that were recorded during the baseline survey (W1). Based on the foregoing, we filtered out the outcomes where the health knowledge was already high at the baseline. For example, 99% of the participants at the baseline acknowledge that fathers should help care for their sick children while 96% of the participants acknowledge that women ideally should wait to have their first baby at 18 years of age or older. We further filtered out the outcomes where the health knowledge was very low at baseline and very low at the endline survey. For example, only 1-2% of participants at the baseline and endline know that difficulty in urinating is a prenatal danger sign, while only 1% of participants at the baseline and endline know that retained placenta is a postpartum danger sign. Following this, we shortlisted 31 knowledge outcomes including two previously unheard riddles. Then, we analyzed the responses of the *targeted* individuals from the 88 randomized targeted villages at the endline survey.

**Statistical analysis**

Given that our focus here is to estimate the knowledge outcomes of intervention at the endline survey (W3) for the *treated* individuals (who got the educational intervention), we used multivariate logistic regression fitted to individual-level data. We estimated the main effect of social interactions on the treated individuals based on the number of ties they have (i.e., we computed an "egocentric reduction" whereby information collected sociocentrically was assigned to each villager) corresponding to the social network built using three name-generated questions (*Who do you spend your free time with?", "Who is your closest friend?",* and *"Who do you discuss personal matters with?"*). Furthermore, to help account for the between-targeted-participants interference within villages, we included the proportion of each ego's ties who also received the specific intervention module. We assume that a respondent's outcome can be affected more by interactions with treated individuals compared to untreated individuals.

In addition, we account for the variability of intervention visits individuals received. In particular, some targeted individuals receive counseling sessions for intervention modules several times. The multiple counseling sessions might act as a direct booster treatment, and we control for this in our model.

We control for a range of demographic variables that may influence knowledge retention and outcomes, including age, gender, marital status, education level, and economic status. Existing research on cognitive aging indicates that memory performance tends to decline from early to late adulthood[28], largely due to a general slowing of cognitive processes and a reduction in attentional resources[29,30,31]. Early research on gender differences in memory tasks has suggested that while neither sex can be said to have better memory per se, the two sexes differ in terms of what type of information they remember best[32]. Women outperform men on memory tasks that require remembering verbally encoded items owing to their advantage in verbal ability, while men's advantage in visuospatial ability favors them in the visual episodic memory performance[33]. These differences have been suggested to stem from gender-specific physiological capabilities, interests, expectations, or some complex interactions of these factors[32]. Females tend to remember information better such as being a responsible mother and taking good care of their children which tapped their specific interests to a greater extent[33]. We also account for marital status since married individuals might have a more vested interest in the information related to mother and child health care practices.

The influence of schooling on cognitive ability is well documented, showing a significant association of low schooling level with lower performance on attention, learning and episodic memory, reaction time, and spatial working memory[34]. Recent studies have shown the influence of formal schooling, particularly literacy acquisition on the growth of memory abilities, specifically on verbal memory[35]. This is theorized to be due to improvements in processing and temporary encoding of spoken language through phoneme awareness[36].

Finally, socioeconomic disparities are associated with individual differences in cognitive development. Studies have found family income is logarithmically associated with brain surface area in children and adolescents[37]. Another line of research has suggested that individuals in low-income families are exposed to chronic stress (food anxiety, neighborhood disadvantages, and parental stress), which is known to have an impact across the brain, specifically on frontal areas, impairing both working memory and long-term memory[38]. Lower income is also associated with greater inflammatory activity (poor diet, fewer healthy behaviors, exposure to pathogens) which in turn impairs cognitive functioning[39].

Last, we accounted for the time gap, estimated as the number of months between the last intervention visit and the endline survey start date to account for forgetting in line with Ebbinghaus' forgetting curves[12].

Significance was defined as $p<0.05$. All analyses were performed using R V.4.2.2.

## RESULTS

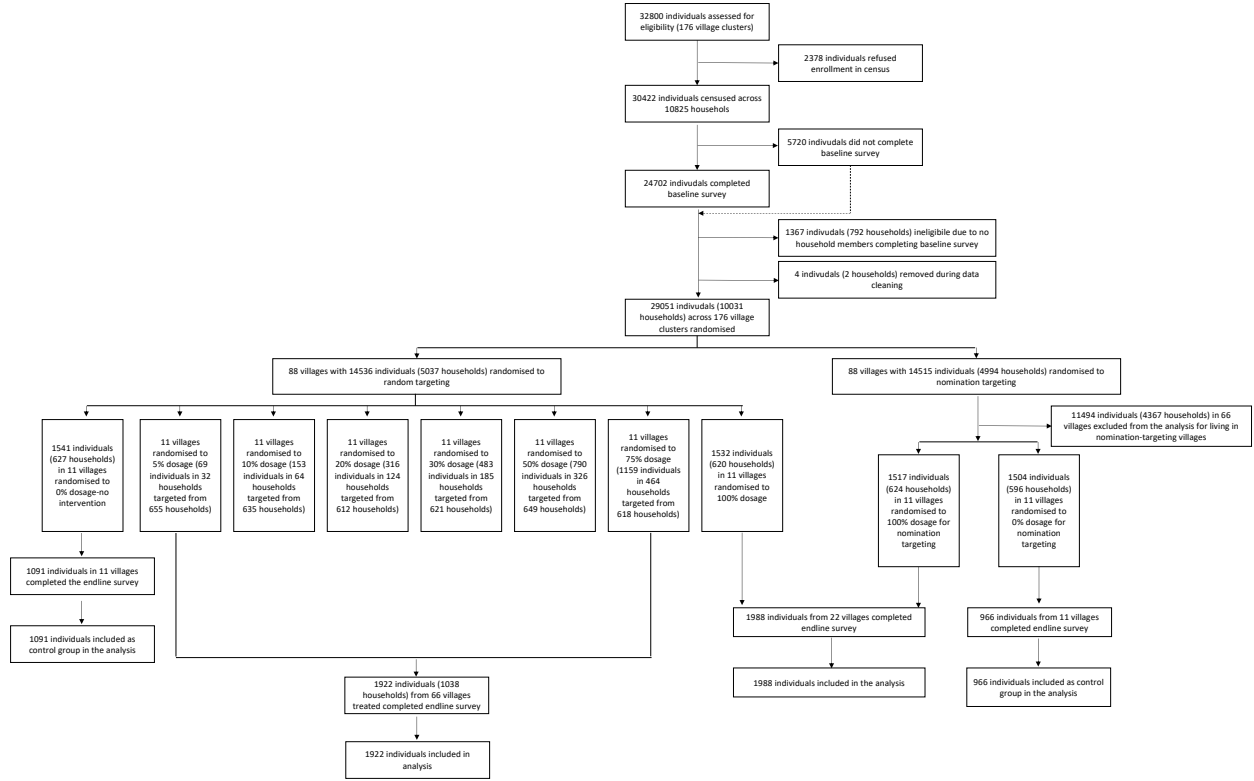

**Fig 1. A population-level cluster randomized controlled trial of maternal and child health intervention**

Participants aged 12 and older were recruited and completed the baseline survey between June 2015 and July 2016. A total of 66 of 176 villages were randomized to the random targeting strategy, in which 11 villages were randomly assigned to treatment dosages or proportion of households targeted, of between 5% and 75%; 22 villages were randomized to 100% targeting; and 22 villages were randomized to control arm with no treated households. Among the treatment arms, a total of 15177 individuals across 5416 households were randomized, 12353 of whom from 5033 households completed the baseline surveys.

Following the 22 months of counseling intervention that began in November 2016, 8250 individuals across 4229 households completed the endline survey (W3) between January 2019 and December 2019, approximately 26 months after intervention delivery began. Of the individuals lost by the endline, 46% moved out of the village, 31% could not be reached, 13% refused to continue in the study, 5% were lost to death, and 5% to other reasons. Participants lost to the endline survey were more likely to be younger, male, single, and have at least a primary level of education.

Across all the 66 villages assigned to random targeting strategy and 22 villages assigned to 100% households targeting, the number of participant households varied between 12 to 193, and the number of individual participants varied between 15 to 315 each. In the randomized targeting villages and 100% targeted villages, a total of 3910 individuals in 2038 households received knowledge intervention visits between November 2016 and August 2018.

Table 1  Characteristics of the population in random-targeting and 100% targeted villages

| Respondent characteristics | Baseline | Endline | Targeted individuals at endline |
|---|---|---|---|
| Individuals | 12353 | 8250 | 3910 |
| Sex | | | |
| Female | 7108 (58 %) | 5321 (65%) | 2506 (64%) |
| Male | 5245 (42 %) | 2929 (35%) | 1404 (36%) |
| Age | 32·8 (17·2) | 38·4 (17·5) | 38·7 (17·5) |
| Education | | | |
| Less than primary | 2616 (21%) | 1954 (24%) | 930 (24%) |
| Primary | 8488 (69%) | 5360 (65%) | 2548 (65%) |
| Secondary or greater | 1249 (10%) | 936 (11%) | 432 (11%) |
| Marital status | | | |
| Single | 4392 (36%) | 2494 (30%) | 1159 (30%) |
| Married or civil union | 7154 (58%) | 5133 (62%) | 2468 (63%) |
| Separated/divorced/widowed | 807 (6%) | 623 (8%) | 283 (7%) |
| Household characteristics | | | |
| Households | 5033 | 4229 | 2038 |
| Wealth index | | | |
| Quintile 1 | 2052 (17%) | 1271 (15%) | 588 (15%) |
| Quintile 2 | 2335 (19%) | 1598 (19%) | 739 (19%) |
| Quintile 3 | 2480 (21%) | 1617 (20%) | 753 (19%) |
| Quintile 4 | 2602 (21%) | 1878 (23%) | 904 (23%) |
| Quintile 5 | 2735 (22%) | 1880 (23%) | 921 (24%) |

In the targeted villages, at baseline, 7108 (58%) participants were women. The mean age of the participants was 32·8 years (SD 17·2); 7154 (58%) of the respondents were married or in a civil union. At the endline survey, among the randomized targeted individuals, the mean age was 38·7 (SD 17·5); 2506 (64%) participants were women; 2468 (63%) of the individuals were married or in a civil union. Descriptive statistics of the randomized targeted villages are reported in Table 1.

The number of friends that individual egos had (an indicator of social interactions) is associated with improvements in the correct knowledge response related to intervention even 20 months after its delivery. We modeled the heterogenous treatment effects of social interaction opportunities by comparing individuals with low (two or fewer friends) versus high (three or more friends) at the endline survey for the treated (from the 88 villages) and the control group (22 villages where no household received any intervention).

Fig. 2. shows four illustrative outcomes corresponding to intervention modules related to pregnancy and postnatal mother care, postnatal newborns care, and knowledge about proper diarrhea treatment among the targeted participants compared to the control group at the endline survey. While the intervention resulted in higher probability of correct response among the targeted individuals, this improvement is significantly higher for individuals with a higher number of friends. For example, while the intervention improved the correct response of knowing the use of folic acid before pregnancy by ~6% for individuals having two friends or less (low opportunity of social interactions), it improves by ~11% for individuals having three or more friends (high opportunity of social interactions). Similarly, the intervention improved the understanding of proper treatment of diarrhea in children five and under by ~8% for individuals having two friends or less, it improves by ~ 18% for individuals having three friends or more. These indicate that the improvements in the knowledge response are several magnitudes higher for more connected individuals.

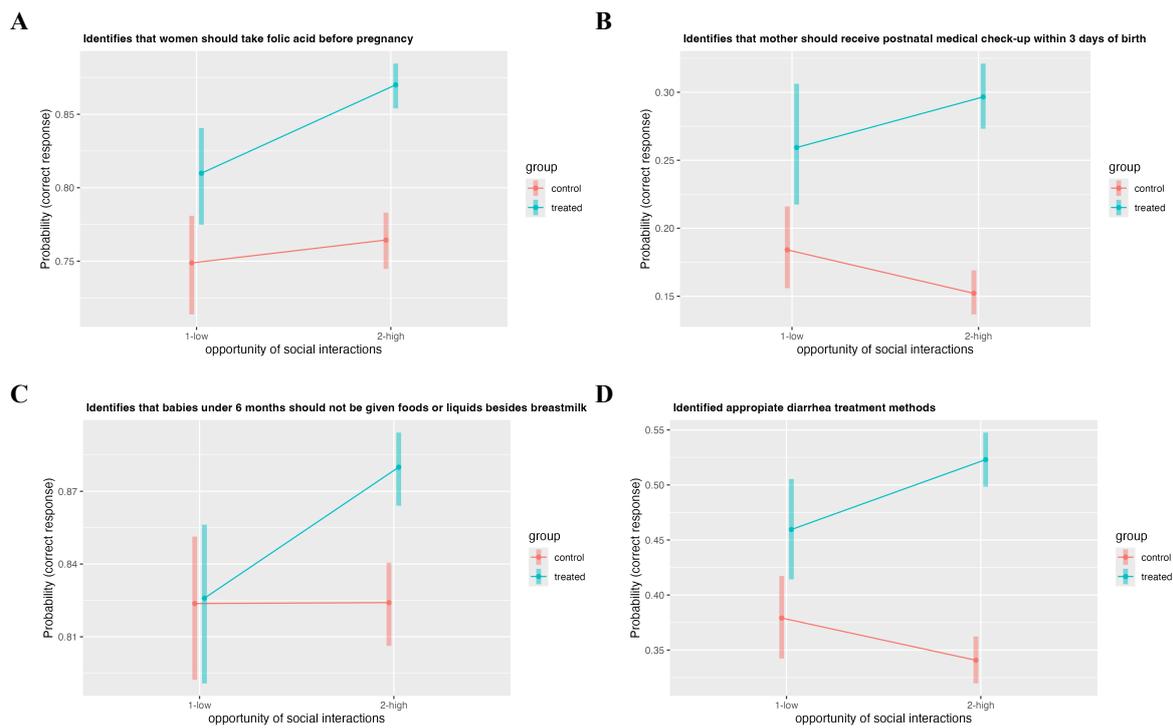

**Fig. 2. Association of the number of friendship ties with knowledge outcomes in treated vs control group**

We estimated the effect size of the friendship ties on the likelihood of correct response for the treated individuals in the focal 88 villages at the endline survey. During the intervention visits, aimed at increasing awareness of health practices among families, community health workers also introduced and taught two previously unheard riddles. The first riddle (riddle 1) asks, "*Dry, dry without a fajero, it falls off quicker, you will see it. What is it?*" (umbilical cord), and the second (riddle 2) states, "*It seems like it is for the roof, but it's not—it's for diarrhea, you tell me, what is it?*" (zinc). We modeled the outcomes of two riddles at the endline survey (W3) for the targeted individuals to observe how friendship ties are associated with the likelihood of correct answer retention to this exogenously introduced and novel information. Our findings, shown in Table 2, indicate that the number of friendship ties is positively associated with the likelihood of providing accurate responses to the riddles. In particular, each additional friendship ties is associated with an average of 30% increase in the probability of a correct response. Moreover, the accuracy of responses to riddles also improves when a higher proportion of an ego's friends (alters) also received the intervention.

**Table 2 Association of friendship ties with riddle knowledge outcomes at endline survey**

|  | Riddle 1 | | Riddle 2 | |
| --- | --- | --- | --- | --- |
|  | effect | p-value♠ | effect | p-value♠ |
| age at survey | -0·742 | <0·000*** | -0·673 | <0·000*** |
| gender male | -0·798 | <0·000*** | -1·316 | <0·000*** |
| edu_level-Primary | 0·508 | 0·001** | 0·431 | 0·004** |
| edu_level-Secondary | 0·911 | 0·009** | 1·031 | 0·003** |
| marital_status-Married | 0·865 | <0·000*** | 0·762 | <0·000*** |
| marital_status-Divorced | 0·409 | 0·144 | 0·262 | 0·328 |
| household_wealth_index2 | 0·204 | 0·390 | 0·151 | 0·535 |
| household_wealth_index3 | 0·267 | 0·251 | -0·062 | 0·796 |
| household_wealth_index4 | 0·060 | 0·760 | 0·233 | 0·340 |
| household_wealth_index5 | 0·410 | 0·096(.) | 0·463 | 0·051(.) |
| month_gap | -0·052 | 0·437 | 0·017 | 0·796 |
| Intervention count | -0·078 | 0·251 | 0·046 | 0·575 |
| Number of friendship ties | 0·245 | <0·000*** | 0·290 | <0·000*** |
| Percent of friends got intervention | 0·043 | 0·498 | 0·178 | 0·005** |

♠ - False discovery rate (FDR) adjusted p-value

Analysing responses across multiple outcomes, we found that the number of friends that targeted individuals have – a proxy for increased opportunities for social interactions – has a significant effect on their likelihood of retaining accurate knowledge at the endline survey related to prenatal care, postnatal mother, and newborn care (see Fig 3).

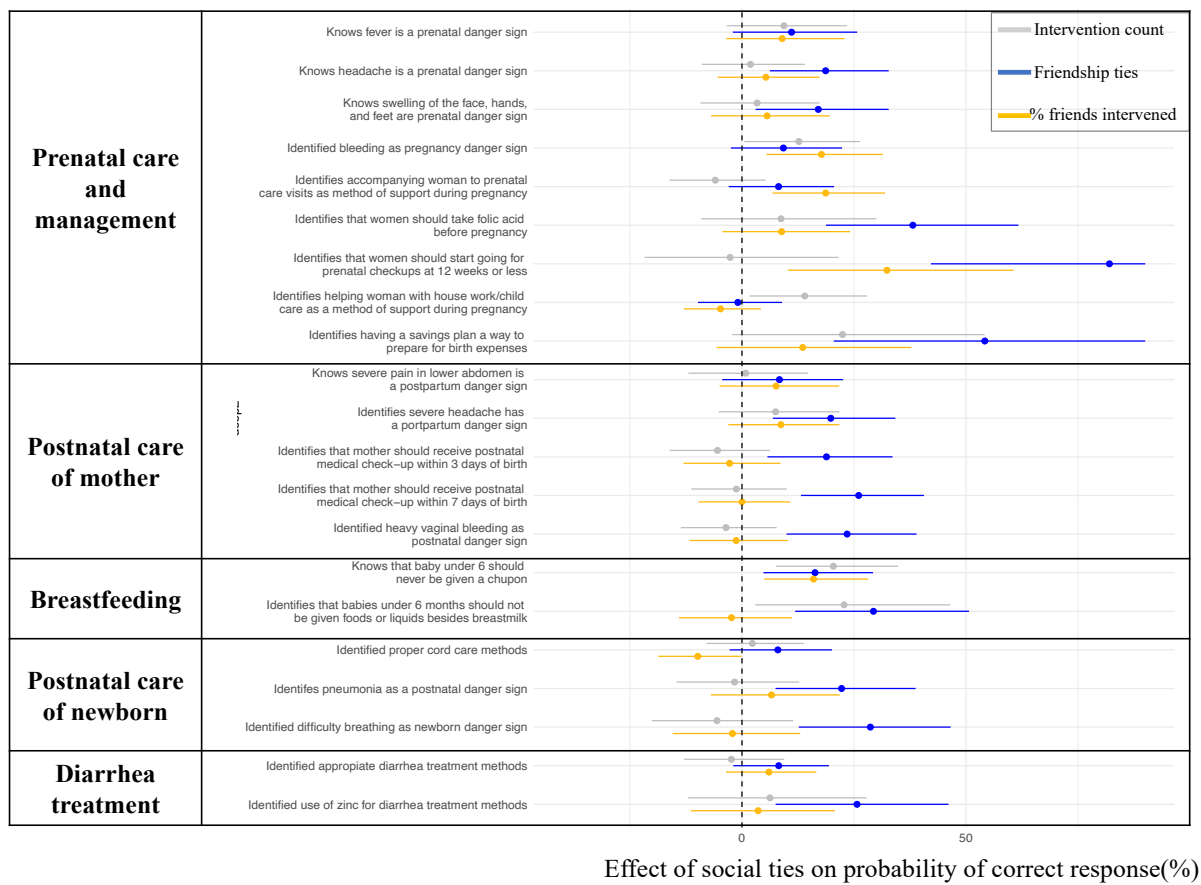

**Fig 3. Significant positive effects of friendship ties on knowledge outcomes at endline survey**

Specifically, an increase in one friendship tie led to a 38·1% (95% CI 18·8%-61·8%, p < 0·000) increase in the likelihood of correctly retaining the knowledge of the use of folic acid before pregnancy for women; a 82·0% (95% CI 42·2%-136·6%, p < 0·000) increase in identifying that women should start going for prenatal checkups at 12 weeks or less during pregnancy; and a 18·7% (95% CI 6·2%-32·7%, p=0·010) increase in the probability of knowing that headache during pregnancy is a danger sign; a 54·2% (95% CI 20·4%-100·6%, p=0·006) increase in identifying having a savings plan is a way to prepare for birth expenses.

The effectiveness of knowledge intervention related to postnatal mother and newborn care is also significantly increased with an increase in friendship ties, leading to a 18·9% (95% CI 5·7%-33·6%, p=0·014) increase in identifying that mother should receive postnatal medical checkup within 3 days of birth; a 26·1% (95% CI 13·2%-40·6%, p < 0·000) increase in identifying that mother should receive postnatal medical checkup within 7 days of birth; a 19.8% (95% CI 6·9%-34·2%, p=0·007) increase in knowing that severe headache is a postpartum danger sign; a 23·5% (95% CI 10·0%-39·0%, p=0·003) increase in knowing that heavy vaginal bleeding is a postpartum danger sign; a 29·4% (95% CI 11·9%-50·7%, p=0·004) increase in knowing that newborn should only be given breastmilk until 6 months; a 28·7% (95% CI 12·7%-46·6%, p=0·001) increase in identifying difficulty in breathing in new-born as danger sign and a 22·2% (95% CI 7·5%-38·8%, p=0·015) increase in identifying that pneumonia is a postnatal danger sign. Furthermore, we found a significant association of friendship ties with the effectiveness of intervention for knowledge related to proper treatment of diarrhea in children five and under.

In addition to the observed association of friendship ties, we also found that the *proportion of friends who were randomly selected to also received the same intervention* as that of the target individual also positively influenced knowledge retention for a few outcomes. For example, an increase in friends who received same intervention increases the likelihood of identifying accompanying woman to prenatal care as support during pregnancy by 18·7% (95% CI 6·8%-31·9%, p=0·011); identifying that bleeding is a prenatal danger sign by 17·7% (95% CI 5·5%-31·5% p=0·008); identifying that women should start going for prenatal checkups at 12 weeks or less during pregnancy by 32·4% (95% CI 10·2%-60·6%, p=0·010); increase in knowing that newborn under the age of 6 months should not be given chupons by 16·0% (95% CI 5·0%-28·2%, p=0·014).

We found no significant effect of friendship ties on knowledge response outcomes related to attitudes towards helping women with housework during pregnancy; awareness of fever as a prenatal danger sign; awareness of feeling pain and weakness as a postpartum danger sign; and acknowledging accompanying woman to prenatal care as support during pregnancy (see online supplementary appendix Table S1).

We performed additional analyses as a check for the robustness of our results, which included building a social network for analysis with additional ties. In particular, since we hypothesize that increased opportunities of social interactions are likely to improve knowledge retention among targetted individuals, we included ties corresponding to two additional name generator questions: whom the respondent gets health advice from, and whom the respondent gives health advice to, in addition to the three friendship ties. We find a similar effect of the number of connections individuals have on the likelihood of them retaining and responding to the knowledge outcomes using this expanded ascertainment of network ties (see online supplementary appendix Table S2 and Table S3).

**DISCUSSION**

An increasing number of real-world interventions are being offered to preemptively or therapeutically protect people against prevailing misbeliefs or malpractices[40]. These interventions have demonstrated positive effects on individuals in fostering correct beliefs or knowledge when measured immediately after treatment. However recent studies have highlighted that such interventions are often only effective for a month before the decay process starts[11]. To counter such decay, memory-enhancing booster interventions have been suggested[18]. However, the scalability of these booster interventions poses challenges, especially under resource-constraint settings.

In this study, we found that being well-connected within a social network might enhance the effectiveness of knowledge interventions. These individuals may be more likely to internalize and retain the information, and reinforced it in others, due to increased opportunities for social interaction where they teach others or learn from them, a mechanism we refer to as "social boosting."

Using data from a randomized controlled trial conducted across 110 villages in Honduras, aimed at delivering knowledge-based interventions on maternal and neonatal health practices, we evaluated the association of having social connections with knowledge retention among targeted individuals, 26 months after the educational intervention began. This study's extended follow-up period enables a robust examination of sustained knowledge retention over time. Our findings indicate a significant positive association between the number of friendship ties an ego has and improved knowledge retention across several key maternal and neonatal health outcomes.

With the increase in one friendship tie, individuals are 38% more likely to know about using folic acid before pregnancy; 82% more likely to acknowledge that pregnant women should start visiting health clinics at 12 weeks or less; 29% more likely to know that newborns less than 6 months should not be fed anything other than breastmilk; and 22% more likely to know pneumonia to be a danger sign for newborns. Furthermore, the effect is enhanced when a larger fraction of an ego's friends got the educational intervention too. With the increase in proportion of ego's friends (alters) intervened upon, targeted individuals are 32% more likely to identify that women should start going for prenatal checkups at 12 weeks or less during pregnancy and 19% more likely to acknowledge that woman should be accompanied to prenatal care visit as a method of support during pregnancy.

These findings underscore the role of social interactions in reinforcing health knowledge interventions over the long term. From a theoretical standpoint, these results are consistent with the generative learning theory and the social presence hypothesis which holds that deeper cognitive processing occurs when learners engage in generative processing leading to improved learning outcomes (although, to be clear, here we only observed *opportunities* to engage, not actual conversations). In particular, possibly by explaining acquired knowledge to others, learners can engage in the cognitive processes of selecting relevant information, organizing it in a coherent structure, and integrating it with prior knowledge. This process promotes the formation of both internal cognitive connections and external communicative linkages, enhancing learning and long-term knowledge retention[19]. Furthermore, the social presence hypothesis posits that heightened perceptions of social presence—such as those experienced when explaining information to others—enhance motivation and elicit physiological arousal[41]. These factors foster the consolidation of working memory and stimulate more intensive cognitive processing, thereby promoting deeper learning outcomes[42].

Alternatively, insights from cognitive neuroscience studies that measure brain activity during cognitive and social processing, offer a potential mechanism through which social interactions may enhance cognitive functioning. These studies highlight that regions of the brains that are activated during social interactions such as explaining to others, are also associated with long term-memory and integrating information from learning materials[43,44]. In a recent fNIRS study, it is shown that groups involved in both social and cognitive processing have enhanced activated brain regions compared to groups that are engaged in only cognitive tasks. Thus, the simultaneous cognitive and social processing that activates while explaining to others facilitates stronger brain patterns, leading to deeper learning and memory outcomes[45].

The enhanced effect of intervened alters on egos' learning that we observe can also be understood through the lens of active and collaborative learning theories[46]. Research on active learning underscores the value of peer discussions in promoting knowledge retention. These discussions, a facet of active learning, promote critical thinking and higher-order, deep learning[47]. When learners acquire new knowledge, they benefit significantly from interactive exchanges with their peers, in which they elaborate, defend, and extend their positions, opinions and beliefs[48]. Such processes promote the resolution of any inconsistencies, stimulate new ideas, and enhance elaboration of more coherent mental models[49]. This deeper cognitive processing not only strengthens knowledge structures, but also counteracts forgetting, thereby facilitating long-term retention of newly acquired information[50].

It is noteworthy that while we observed a generally positive association of friendship ties with knowledge outcomes across several outcomes, the extent and nature of this association may vary depending on the knowledge possibly being transmitted – such as the complexity of the knowledge or the prevalence of the long-standing beliefs. For instance, in the case of the outcome assessing whether individuals recognize providing support to pregnant women (such as assisting with household work or childcare) as an appropriate form of help, our analysis revealed a non-significant negative association with the number of friendship ties. This finding suggests that normative beliefs, particularly those tied to entrenched gender roles, may be resistant to change either due to reduced interpersonal dissemination by targetted individuals or the reinforcement of prevailing norms through social feedback. Interestingly, for this same outcome, we found a significant positive effect of intervention visits by health agents, indicating that repeated exposure (direct booster treatment) to targeted messaging can have a durable impact, in the face of resistant cultural norms.

This work has limitations. First, the data used in this study is collected through RCT that focuses on one country and an intervention specific to maternal and child health. Future studies could study the social booster effects of correcting misbeliefs (misinformation) with respect to political conspiracy theories, scientific myths around epidemics, or climate change perspectives. Second, the current study used name generator questions to capture social network interactions. It will be interesting to analyze more granular social interactions, e.g., by asking how many times targeted individuals actually discussed the new knowledge they acquired with their friends, who they discussed with, and so on. These will help establish more causal relationships of the underlying mechanism. Third, the current work studies information flow in a face-to-face network, and online social networks may be different.

The social reinforcement of learning is a key means by which learning transpires in the first place, and a key means by which forgetting is avoided. When a subject is surrounded by others with whom they might discuss novel information, they may be more likely to retain, and not just spread, that information themselves.

**Contributors:**

V.K. conceptualized the study, conducted formal analysis, and wrote the original draft. N.A.C. supervised the data collection. Both authors contributed to the interpretation of the results, editing, and revising of the final manuscript. Both authors had full access to all the data in the study and had final responsibility for the decision to submit for publication.

**Declaration of interests**

We declare no competing interests.

**Data sharing**

Compliant with our privacy and confidentiality assurances to our research participants and with other legal obligations, data will be made available on our secure server, subject to data release provisions in force at Yale and the Yale Institute for Network Science (or successor entities) at the time of release. Access to data requires proof of IRB approval and human participants certification. Contact nicholas.christakis@yale.edu for inquiries regarding the data.

**Acknowledgments**

We acknowledge support from Grant OPP1098684 from the Bill and Melinda Gates Foundation. VK acknowledges support from Grant No. 217720 from Swiss National Science Foundation (SNSF).

# REFERENCES

1. Pennycook, G., Cannon, T. D. & Rand, D. G. Prior exposure increases perceived accuracy of fake news. *Journal of Experimental Psychology: General* **147**, 1865–1880 (2018).

2. Cook, J., Lewandowsky, S. & Ecker, U. K. H. Neutralizing misinformation through inoculation: Exposing misleading argumentation techniques reduces their influence. *PLoS ONE* **12**, e0175799 (2017).

3. Hassan, A. & Barber, S. J. The effects of repetition frequency on the illusory truth effect. *Cogn. Research* **6**, 38 (2021).

4. Lewandowsky, S., Oreskes, N., Risbey, J. S., Newell, B. R. & Smithson, M. Seepage: Climate change denial and its effect on the scientific community. *Global Environmental Change* **33**, 1–13 (2015).

5. Loomba, S., De Figueiredo, A., Piatek, S. J., De Graaf, K. & Larson, H. J. Measuring the impact of COVID-19 vaccine misinformation on vaccination intent in the UK and USA. *Nat Hum Behav* **5**, 337–348 (2021).

6. Ahmed, W., Vidal-Aballl, J., Downing, J. & López Seguí, F. COVID-19 and the 5G Conspiracy Theory: Social Network Analysis of Twitter Data. *J Med Internet Res* **22**, e19458 (2020).

7. Chan, M. S., Jones, C. R., Hall Jamieson, K. & Albarracín, D. Debunking: A Meta-Analysis of the Psychological Efficacy of Messages Countering Misinformation. *Psychol Sci* **28**, 1531–1546 (2017).

8. Compton, J. A. & Pfau, M. Inoculation Theory of Resistance to Influence at Maturity: Recent Progress In Theory Development and Application and Suggestions for Future Research. *Annals of the International Communication Association* **29**, 97–146 (2005).

9. Ivanov, B. Inoculation Theory Applied in Health and Risk Messaging. in *Oxford Research Encyclopedia of Communication* (Oxford University Press, 2017). doi:10.1093/acrefore/9780190228613.013.254.

10. Banas, J. A. & Rains, S. A. A Meta-Analysis of Research on Inoculation Theory. *Communication Monographs* **77**, 281–311 (2010).

11. Ivanov, B., Parker, K. A. & Dillingham, L. L. Testing the Limits of Inoculation-Generated Resistance. *Western Journal of Communication* **82**, 648–665 (2018).

12. Rubin, D. C. & Wenzel, A. E. One hundred years of forgetting: A quantitative description of retention. *Psychological Review* **103**, 734–760 (1996).

13. Zerback, T., Töpfl, F. & Knöpfle, M. The disconcerting potential of online disinformation: Persuasive effects of astroturfing comments and three strategies for inoculation against them. *New Media & Society* **23**, 1080–1098 (2021).

14. Ecker, U. K. H., Lewandowsky, S. & Tang, D. T. W. Explicit warnings reduce but do not eliminate the continued influence of misinformation. *Mem Cogn* **38**, 1087–1100 (2010).

15. Ecker, U. K. H., Lewandowsky, S., Fenton, O. & Martin, K. Do people keep believing because they want to? Preexisting attitudes and the continued influence of misinformation. *Mem Cogn* **42**, 292–304 (2014).

16. Hardt, O., Nader, K. & Nadel, L. Decay happens: the role of active forgetting in memory. *Trends in Cognitive Sciences* **17**, 111–120 (2013).

17. Murre, J. M. J. & Dros, J. Replication and Analysis of Ebbinghaus' Forgetting Curve. *PLoS ONE* **10**, e0120644 (2015).

18. Maertens, R. *et al.* Psychological booster shots targeting memory increase long-term resistance against misinformation. *Nat Commun* **16**, 2062 (2025).

19. Fiorella, L. & Mayer, R. E. The relative benefits of learning by teaching and teaching expectancy. *Contemporary Educational Psychology* **38**, 281–288 (2013).

20. Cohen, J. Theoretical considerations of peer tutoring. *Psychol. Schs.* **23**, 175–186 (1986).

21. King, A., Staffieri, A. & Adelgais, A. Mutual peer tutoring: Effects of structuring tutorial interaction to scaffold peer learning. *Journal of Educational Psychology* **90**, 134–152 (1998).

22. Shakya, H. B. *et al.* Intimate partner violence norms cluster within households: an observational social network study in rural Honduras. *BMC Public Health* **16**, (2016).

23. Airoldi, E. M. & Christakis, N. A. Induction of social contagion for diverse outcomes in structured experiments in isolated villages. *Science* **384**, eadi5147 (2024).

24. Lungeanu, A. *et al.* Using Trellis software to enhance high-quality large-scale network data collection in the field. *Social Networks* **66**, 171–184 (2021).

25. Shakya, H. B. *et al.* Exploiting social influence to magnify population-level behaviour change in maternal and child health: study protocol for a randomised controlled trial of network targeting algorithms in rural Honduras. *BMJ Open* **7**, e012996 (2017).

26. *World Vision International. Timed and Targeted Counseling (ttC)*. https://www.wvi.org/health/timed-and-targeted-counseling-ttc (2016).

27. *Health Communication Capacity Collaborative. The P Process: Five Steps to Strategic Communication*. https://www.healthcommcapacity.org/wp-content/uploads/2014/04/P-Process-Brochure.pdf (2013).

28. Grady, C. Changes in memory processing with age. *Current Opinion in Neurobiology* **10**, 224–231 (2000).

29. Craik, F. I. M. & Byrd, M. Aging and Cognitive Deficits. in *Aging and Cognitive Processes* (eds. Craik, F. I. M. & Trehub, S.) 191–211 (Springer US, Boston, MA, 1982). doi:10.1007/978-1-4684-4178-9_11.

30. Craik, F. I. M., Morris, L. W., Morris, R. G. & Loewen, E. R. Relations between source amnesia and frontal lobe functioning in older adults. *Psychology and Aging* **5**, 148–151 (1990).

31. Naveh-Benjamin, M. Adult age differences in memory performance: Tests of an associative deficit hypothesis. *Journal of Experimental Psychology: Learning, Memory, and Cognition* **26**, 1170–1187 (2000).
32. Loftus, E. F., Banaji, M. R., Schooler, J. W. & Foster, R. Who remembers what?: Gender differences in memory. *Michigan Quarterly Review* **26**, 64–85 (1987).
33. Pauls, F., Petermann, F. & Lepach, A. C. Gender differences in episodic memory and visual working memory including the effects of age. *Memory* **21**, 857–874 (2013).
34. Bento-Torres, N. V. O. *et al.* Influence of schooling and age on cognitive performance in healthy older adults. *Braz J Med Biol Res* **50**, e5892 (2017).
35. Kolinsky, R. *et al.* The influence of age, schooling, literacy, and socioeconomic status on serial-order memory. *J Cult Cogn Sci* **4**, 343–365 (2020).
36. Demoulin, C. & Kolinsky, R. Does learning to read shape verbal working memory? *Psychon Bull Rev* **23**, 703–722 (2016).
37. Noble, K. G. *et al.* Family income, parental education and brain structure in children and adolescents. *Nat Neurosci* **18**, 773–778 (2015).
38. Shields, G. S. *et al.* Recent life stress exposure is associated with poorer long-term memory, working memory, and self-reported memory. *Stress* **20**, 598–607 (2017).
39. Shields, G. S., Moons, W. G. & Slavich, G. M. Inflammation, Self-Regulation, and Health: An Immunologic Model of Self-Regulatory Failure. *Perspect Psychol Sci* **12**, 588–612 (2017).
40. Lewandowsky, S. & Van Der Linden, S. Countering Misinformation and Fake News Through Inoculation and Prebunking. *European Review of Social Psychology* **32**, 348–384 (2021).
41. Hoogerheide, V., Deijkers, L., Loyens, S. M. M., Heijltjes, A. & Van Gog, T. Gaining from explaining: Learning improves from explaining to fictitious others on video, not from writing to them. *Contemporary Educational Psychology* **44–45**, 95–106 (2016).
42. Fiorella, L. & Mayer, R. E. Eight Ways to Promote Generative Learning. *Educ Psychol Rev* **28**, 717–741 (2016).
43. Jakobs, O. *et al.* Across-study and within-subject functional connectivity of a right temporo-parietal junction subregion involved in stimulus–context integration. *NeuroImage* **60**, 2389–2398 (2012).
44. Decety, J. & Lamm, C. The Role of the Right Temporoparietal Junction in Social Interaction: How Low-Level Computational Processes Contribute to Meta-Cognition. *Neuroscientist* **13**, 580–593 (2007).
45. Zhu, W., Wang, F., Mayer, R. E. & Liu, T. Effects of explaining a science lesson to others or to oneself: A cognitive neuroscience approach. *Learning and Instruction* **91**, 101897 (2024).
46. Van Blankenstein, F. M., Dolmans, D. H. J. M., Van Der Vleuten, C. P. M. & Schmidt, H. G. Which cognitive processes support learning during small-group discussion? The role of providing explanations and listening to others. *Instr Sci* **39**, 189–204 (2011).

47. Pollock, P. H., Hamann, K. & Wilson, B. M. Learning Through Discussions: Comparing the Benefits of Small-Group and Large-Class Settings. *Journal of Political Science Education* **7**, 48–64 (2011).
48. Garside, C. Look who's talking: A comparison of lecture and group discussion teaching strategies in developing critical thinking skills. *Communication Education* **45**, 212–227 (1996).
49. Webb, N. M. Peer interaction and learning in small groups. *International Journal of Educational Research* **13**, 21–39 (1989).
50. Reder, L. M. The Role of Elaboration in the Comprehension and Retention of Prose: A Critical Review. *Review of Educational Research* **50**, 5–53 (1980).

# APPENDIX

**Table S1. The effect size of friend ties (using 'personal-private matter', 'spent free time', and 'closest friend' name generator)**

|  | N | Friendship ties effect size (95% CI) | p-value♦ | Prop. Friends Intervene effect size (95% CI) | p-value♦ |
|---|---|---|---|---|---|
| **Intervention knowledge** | | | | | |
| Correctly answered proper cord care riddle | 1487 | 0·25*** (0·12 to 0·37) | < 0·000 | 0·04 (-0·07 to 0·16) | 0·498 |
| Correctly answered diarrhea treatment with zinc riddle | 1633 | 0·29*** (0·17 to 0·41) | < 0·000 | 0·18** (0·07 to 0·29) | 0·005 |
| **Prenatal Care** | | | | | |
| Identifies that women should take folic acid before pregnancy | 2257 | 0·32*** (0·17 to 0·48) | < 0·000 | 0·08 (-0·04 to 0·22) | 0·320 |
| Identifies that women should seek prenatal care first 12 weeks of pregnancy | 1890 | 0·60*** (0·35 to 0·86) | < 0·000 | 0·28** (0·10 to 0·47) | 0·010 |
| Identifies accompanying woman to prenatal care visits as method of support during pregnancy | 1548 | 0·08 (-0·03 to 0·19) | 0·465 | 0·17* (0·07 to 0·28) | 0·011 |
| Identifies helping woman with house work/child care as method of support during pregnancy | 2069 | -0·01 (-0·10 to 0·09) | 0·849 | -0·05 (-0·14 to 0·04) | 0·460 |
| Identified fever as pregnancy d.s. | 1548 | 0·10 (-0·02 to 0·23) | 0·197 | 0·09 (-0·04 to 0·21) | 0·205 |
| Identified headache as pregnancy d.s. | 1548 | 0·17** (0·06 to 0·28) | 0·010 | 0·05 (-0·06 to 0·16) | 0·641 |
| Identified swelling of face/hands/feet as pregnancy d.s. | 1548 | 0·16* (0·03 to 0·28) | 0·044 | 0·05 (-0·07 to 0·18) | 0·564 |
| Identified bleeding as pregnancy d.s. | 1548 | 0·09 (-0·03 to 0·20) | 0·172 | 0·16** (0·05 to 0·27) | 0·008 |
| Identified dizziness as pregnancy d.s. | 1548 | 0·07 (-0·04 to 0·18) | 0·407 | 0·07 (-0·04 to 0·17) | 0·407 |
| **Prenatal Management** | | | | | |
| Identifies having a savings plan as method of preparing for birth expenses | 1897 | 0·43** (0·19 to 0·70) | 0·006 | 0·13 (-0·06 to 0·32) | 0·280 |
| **Postnatal Care for mother** | | | | | |
| Identifies that mother should receive postnatal medical check-up within 3 days of birth | 1690 | 0·17* (0·06 to 0·29) | 0·014 | -0·03 (-0·14 to 0·08) | 0·717 |
| Identifies that mother should receive postnatal medical check-up within 7 days of birth | 1690 | 0·23*** (0·12 to 0·34) | < 0·000 | 0·00 (-0·10 to 0·10) | 0·998 |
| Identified severe headache as postpartum d.s. | 1788 | 0·18** (0·07 to 0·29) | 0·007 | 0·08 (-0·03 to 0·20) | 0·283 |
| Identified heavy vaginal bleeding as postpartum d.s. | 1420 | 0·21** (0·09 to 0·33) | 0·003 | -0·01 (-0·12 to 0·10) | 0·961 |
| Identified severe lower abdomen pain as postpartum d.s. | 1548 | 0·08 (-0·05 to 0·20) | 0·512 | 0·07 (-0·05 to 0·20) | 0·525 |
| Identified fever as postnatal d.s. | 1788 | 0·10 (-0·01 to 0·20) | 0·114 | 0·09 (-0·01 to 0·20) | 0·114 |
| Identified weakness or fainting as postnatal d.s. | 1723 | 0·11 (-0·02 to 0·23) | 0·426 | -0·05 (-0·19 to 0·08) | 0·640 |
| **Breastfeeding** | | | | | |
| Identifies that newborns should only be given breast milk during first 6 months | 2168 | 0·26** (0·11 to 0·41) | 0·004 | -0·02 (-0·15 to 0·11) | 0·771 |
| Believes newborns should not be given chupón during first 6 months | 1849 | 0·15* (0·05 to 0·26) | 0·014 | 0·15* (0·05 to 0·25) | 0·014 |
| Identifies that newborn should be breastfed immediately after birth | 1448 | 0·07 (-0·11 to 0·25) | 0·584 | 0·09 (-0·07 to 0·27) | 0·424 |
| **Postnatal care of newborn** | | | | | |
| Identified proper cord care methods | 1716 | 0·08 (-0·03 to 0·18) | 0·380 | -0·10 (-0·21 to 0·00) | 0·142 |
| Identified pneumonia as newborn d.s. | 1716 | 0·20* (0·07 to 0·33) | 0·015 | 0·06 (-0·07 to 0·20) | 0·903 |
| Identified vomiting as newborn d.s. | 2113 | 0·03 (-0·11 to 0·15) | 0·894 | 0·14 (0·01 to 0·27) | 0·110 |
| Identified difficulty breathing as newborn d.s. | 1951 | 0·25** (0·12 to 0·38) | 0·001 | -0·02 (-0·17 to 0·12) | 0·828 |
| Identified fever as newborn d.s. | 1951 | 0·01 (-0·09 to 0·11) | 0·949 | 0·00 (-0·10 to 0·09) | 0·949 |
| Identified diarrhea as newborn d.s. | 1931 | 0·01 (-0·09 to 0·11) | 0·908 | 0·34 (-0·02 to 0·18) | 0·184 |
| Identifies that newborn should be bathed >24 hours after birth | 1883 | -0·10 (-0·26 to 0·05) | 0·656 | -0·05 (-0·19 to 0·09) | 0·656 |
| **Diarrhea Management** | | | | | |
| Identified appropriate diarrhea treatment methods | 1931 | 0·08 (-0·02 to 0·18) | 0·290 | 0·06 (-0·4 to 0·15) | 0·340 |
| Identified use of zinc for diarrhea treatment methods | 1931 | 0·23* (0·07 to 0·38) | 0·013 | 0·04 (-0·12 to 0·19) | 0·816 |

Φ - number of friendship ties using name generators: (i) personal-private matters; (ii) spent free time; (iii) closest friend

♠ - False discovery rate (FDR) adjusted p-value

**Table S2. The effect size of friend ties (using 'personal-private matter', 'spent free time', 'closest friend', 'give health advice to' and 'get health advice from' name generator)**

| | N | Friendship ties effect size (95% CI) | p-value♠ | Prop. Friends Intervene effect size (95% CI) | p-value♠ |
|---|---|---|---|---|---|
| **Intervention knowledge** | | | | | |
| Correctly answered proper cord care riddle | 1487 | 0·24*** (0·12 to 0·36) | < 0·000 | 0·04 (-0·07 to 0·16) | 0·494 |
| Correctly answered diarrhea treatment with zinc riddle | 1633 | 0·33*** (0·21 to 0·45) | < 0·000 | 0·18** (0·07 to 0·29) | 0·005 |
| **Prenatal Care** | | | | | |
| Identifies that women should take folic acid before pregnancy | 2257 | 0·34*** (0·19 to 0·50) | < 0·000 | 0·09 (-0·04 to 0·22) | 0·326 |
| Identifies that women should seek prenatal care first 12 weeks of pregnancy | 1890 | 0·70*** (0·45 to 0·97) | < 0·000 | 0·28** (0·10 to 0·47) | 0·010 |
| Identifies accompanying woman to prenatal care visits as method of support during pregnancy | 1548 | 0·09 (-0·02 to 0·20) | 0·316 | 0·17* (0·07 to 0·28) | 0·011 |
| Identifies helping woman with house work/child care as method of support during pregnancy | 2069 | -0·01 (-0·11 to 0·08) | 0·777 | -0·05 (-0·14 to 0·04) | 0·455 |
| Identified fever as pregnancy d.s. | 1548 | 0·09 (-0·03 to 0·22) | 0·209 | 0·09 (-0·03 to 0·21) | 0·209 |
| Identified headache as pregnancy d.s. | 1548 | 0·20** (0·09 to 0·31) | 0·002 | 0·05 (-0·06 to 0·16) | 0·636 |
| Identified swelling of face/hands/feet as pregnancy d.s. | 1548 | 0·20** (0·08 to 0·33) | 0·005 | 0·05 (-0·07 to 0·18) | 0·617 |
| Identified bleeding as pregnancy d.s. | 1548 | 0·12(.) (0·01 to 0·24) | 0·06 | 0·16** (0·05 to 0·27) | 0·008 |
| Identified dizziness as pregnancy d.s. | 1548 | 0·07 (-0·04 to 0·18) | 0·40 | 0·07 (-0·04 to 0·17) | 0·401 |
| **Prenatal Management** | | | | | |
| Identifies having a savings plan as method of preparing for birth expenses | 1897 | 0·50** (0·25 to 0·77) | 0·001 | 0·12 (-0·06 to 0·32) | 0·296 |
| **Postnatal Care for mother** | | | | | |
| Identifies that mother should receive postnatal medical check-up within 3 days of birth | 1690 | 0·16* (0·04 to 0·27) | 0·029 | -0·03 (-0·14 to 0·08) | 0·733 |
| Identifies that mother should receive postnatal medical check-up within 7 days of birth | 1690 | 0·22*** (0·12 to 0·33) | < 0·000 | 0·00 (-0·10 to 0·10) | 0·977 |
| Identified severe headache as postpartum d.s. | 1788 | 0·19** (0·08 to 0·30) | 0·005 | 0·08 (-0·03 to 0·20) | 0·275 |
| Identified heavy vaginal bleeding as postpartum d.s. | 1420 | 0·21** (0·09 to 0·33) | 0·003 | -0·01 (-0·12 to 0·10) | 0·992 |
| Identified severe lower abdomen pain as postpartum d.s. | 1548 | 0·09 (-0·04 to 0·21) | 0·429 | 0·07 (-0·05 to 0·20) | 0·521 |
| Identified fever as postnatal d.s. | 1788 | 0·33 (-0·01 to 0·20) | 0·111 | 0·10 (-0·01 to 0·20) | 0·111 |
| Identified weakness or fainting as postnatal d.s. | 1723 | 0·17 (-0·02 to 0·23) | 0·421 | -0·05 (-0·19 to 0·08) | 0·648 |
| **Breastfeeding** | | | | | |
| Identifies that newborns should only be given breast milk during first 6 months | 2168 | 0·28** (0·13 to 0·43) | 0·001 | -0·02 (-0·15 to 0·11) | 0·789 |
| Believes newborns should not be given chupón during first 6 months | 1849 | 0·17** (0·06 to 0·27) | 0·007 | 0·15* (0·05 to 0·25) | 0·011 |
| Identifies that newborn should be breastfed immediately after birth | 1448 | 0·09 (-0·08 to 0·28) | 0·410 | 0·09 (-0·07 to 0·27) | 0·410 |
| **Postnatal care of newborn** | | | | | |
| Identified proper cord care methods | 1716 | 0·08 (-0·02 to 0·19) | 0·330 | -0·10 (-0·21 to 0·00) | 0·141 |
| Identified pneumonia as newborn d.s. | 1716 | 0·27*** (0·14 to 0·39) | < 0·000 | 0·06 (-0·07 to 0·20) | 0·848 |
| Identified vomiting as newborn d.s. | 2113 | 0·13 (-0·15 to 0·12) | 0·869 | 0·14 (0·02 to 0·27) | 0·105 |
| Identified difficulty breathing as newborn d.s. | 1951 | 0·12 | < 0·000 | -0·02 | 0·828 |
| | | (0·14 to 0·40) | | (-0·17 to 0·12) | |
| Identified fever as newborn d.s. | 1951 | 0·58 (-0·08 to 0·12) | 0·801 | 0·00 (-0·10 to 0·09) | 0·942 |
| Identified diarrhea as newborn d.s. | 1931 | 0·34 (-0·09 to 0·11) | 0·906 | 0·08 (-0·02 to 0·18) | 0·184 |
| Identifies that newborn should be bathed >24 hours after birth | 1883 | -0·10 (-0·26 to 0·05) | 0·652 | -0·05 (-0·19 to 0·09) | 0·652 |
| **Diarrhea Management** | | | | | |
| Identified appropriate diarrhea treatment methods | 1931 | 0·11(.) (0·01 to 0·21) | 0·094 | 0·06 (-0·04 to 0·15) | 0·429 |
| Identified use of zinc for diarrhea treatment methods | 1931 | 0·25** (0·09 to 0·40) | 0·006 | 0·04 (-0·12 to 0·19) | 0·807 |

Φ - number of friendship ties using name generators: (i) personal-private matters; (ii) spent free time; (iii) closest friend (iv) give health advice to (v) get health advice from

♠ - False discovery rate (FDR) adjusted p-value

**Table S3. The effect size of friend ties (using 'give health advice to' and 'get health advice from' name generator)**

| | N | Friendship ties effect size (95% CI) | p-value♠ | Prop. Friends Intervene effect size (95% CI) | p-value♠ |
|---|---|---|---|---|---|
| **Intervention knowledge** | | | | | |
| Correctly answered proper cord care riddle | 1487 | 0·08 (-0·04 to 0·19) | 0·248 | 0·05 (-0·07 to 0·16) | 0·426 |
| Correctly answered diarrhea treatment with zinc riddle | 1633 | 0·31*** (0.2 to 0.43) | < 0·000 | 0·18** (0·06 to 0·29) | 0·005 |
| **Prenatal Care** | | | | | |
| Identifies that women should take folic acid before pregnancy | 2257 | 0·29*** (0·15 to 0·44) | < 0·000 | 0·08 (-0·05 to 0·22) | 0·310 |
| Identifies that women should seek prenatal care first 12 weeks of pregnancy | 1890 | 0·69*** (0·44 to 0·97) | < 0·000 | 0·31** (0·12 to 0·51) | 0·008 |
| Identifies accompanying woman to prenatal care visits as method of support during pregnancy | 1548 | 0·07 (-0·03 to 0·18) | 0·546 | 0·17* (0·06 to 0·28) | 0·012 |
| Identifies helping woman with house work/child care as method of support during pregnancy | 2069 | 0·05 (-0·04 to 0·14) | 0·441 | -0·05 (-0·14 to 0·04) | 0·441 |
| Identified fever as pregnancy d.s. | 1548 | 0·08 (-0·04 to 0·20) | 0·242 | 0·09 (-0·04 to 0·21) | 0·242 |
| Identified headache as pregnancy d.s. | 1548 | 0·20** (0·09 to 0·31) | 0·002 | 0·05 (-0·06 to 0·16) | 0·725 |
| Identified swelling of face/hands/feet as pregnancy d.s. | 1548 | 0·26*** (0·14 to 0·38) | < 0·000 | 0·05 (-0·08 to 0·18) | 0·592 |
| Identified bleeding as pregnancy d.s. | 1548 | 0·18** (0·07 to 0·29) | 0·003 | 0·16** (0·05 to 0·27) | 0·008 |
| Identified dizziness as pregnancy d.s. | 1548 | 0·06 (-0·05 to 0·16) | 0·539 | 0·07 (-0·04 to 0·17) | 0·220 |
| **Prenatal Management** | | | | | |
| Identifies having a savings plan as method of preparing for birth expenses | 1897 | 0·47** (0·24 to 0·72) | 0·001 | 0·13 (-0·06 to 0·33) | 0·243 |
| **Postnatal Care for mother** | | | | | |
| Identifies that mother should receive postnatal medical check-up within 3 days of birth | 1690 | 0·04 (-0·08 to 0·15) | 0·631 | -0·02 (-0·13 to 0·09) | 0·712 |
| Identifies that mother should receive postnatal medical check-up within 7 days of birth | 1690 | 0·07 (-0·03 to 0·18) | 0·335 | 0·00 (-0·10 to 0·10) | 0·961 |
| Identified severe headache as postpartum d.s. | 1788 | 0·15* (0·04 to 0·26) | 0·021 | 0·08 (-0·03 to 0·19) | 0·293 |
| Identified heavy vaginal bleeding as postpartum d.s. | 1420 | 0·17 * (0·05 to 0·28) | 0·020 | -0·02 (-0·13 to 0·10) | 0·924 |
| Identified severe lower abdomen pain as postpartum d.s. | 1548 | 0·11 (0·00 to 0·23) | 0·170 | 0·07 (-0·05 to 0·19) | 0·565 |
| Identified fever as postnatal d.s. | 1788 | 0·11(.) (0·01 to 0·21) | 0·070 | 0·09 (-0·01 to 0·20) | 0·108 |
| Identified weakness or fainting as postnatal d.s. | 1723 | 0·12 (-0·01 to 0·24) | 0·241 | -0·05 (-0·19 to 0·08) | 0·696 |
| **Breastfeeding** | | | | | |
| Identifies that newborns should only be given breast milk during first 6 months | 2168 | 0·26** (0·11 to 0·41) | 0·003 | -0·03 (-0·16 to 0·11) | 0·742 |
| Believes newborns should not be given chupón during first 6 months | 1849 | 0·12* (0·02 to 0·22) | 0·039 | 0·15** (0·05 to 0·25) | 0·012 |
| Identifies that newborn should be breastfed immediately after birth | 1448 | 0·12 (-0·05 to 0·30) | 0·315 | 0·09 (-0·08 to 0·27) | 0·251 |
| **Postnatal care of newborn** | | | | | |
| Identified proper cord care methods | 1716 | 0·05 (-0·05 to 0·15) | 0·761 | -0·10 (-0·21 to 0·00) | 0·150 |
| Identified pneumonia as newborn d.s. | 1716 | 0·31*** (0·19 to 0·43) | < 0·000 | 0·06 (-0·08 to 0·19) | 0·837 |
| Identified vomiting as newborn d.s. | 2113 | -0·04 (-0·17 to 0·09) | 0·861 | 0·14 (0·02 to 0·27) | 0·101 |
| Identified difficulty breathing as newborn d.s. | 1951 | 0·26*** (0·13 to 0·39) | < 0·000 | -0·02 (-0·17 to 0·12) | 0·771 |
| Identified fever as newborn d.s. | 1951 | 0·14* (0·04 to 0·23) | 0·028 | -0·01 (-0·10 to 0·09) | 0·927 |
| Identified diarrhea as newborn d.s. | 1931 | 0·08 (-0·02 to 0·17) | 0·182 | 0·08 (-0·02 to 0·18) | 0·182 |
| Identifies that newborn should be bathed >24 hours after birth | 1883 | -0·05 (-0·20 to 0·09) | 0·590 | -0·05 (-0·19 to 0·09) | 0·590 |
| **Diarrhea Management** | | | | | |
| Identified appropriate diarrhea treatment methods | 1931 | 0·10 (0·01 to 0·20) | 0·103 | 0·06 (-0·04 to 0·15) | 0·384 |
| Identified use of zinc for diarrhea treatment methods | 1931 | 0·20* (0·05 to 0·34) | 0·024 | 0·04 (-0·12 to 0·19) | 0·806 |

Φ - number of friendship ties using name generators: (i) give health advice to (ii) get health advice from

♠ - False discovery rate (FDR) adjusted p-value